\pdfoutput=1
\RequirePackage{ifpdf}
\ifpdf 
\documentclass[pdftex]{sigma}
\else
\documentclass{sigma}
\fi

\numberwithin{equation}{section}

\begin{document}

\allowdisplaybreaks

\newcommand{\arXivNumber}{1612.00348}

\renewcommand{\thefootnote}{}

\renewcommand{\PaperNumber}{012}

\FirstPageHeading

\ShortArticleName{Irregular Conformal States and Spectral Curve: Irregular Matrix Model Approach}

\ArticleName{Irregular Conformal States and Spectral Curve:\\ Irregular Matrix Model Approach\footnote{This paper is a~contribution to the Special Issue on Combinatorics of Moduli Spaces: Integrability, Cohomo\-logy, Quantisation, and Beyond. The full collection is available at \href{http://www.emis.de/journals/SIGMA/moduli-spaces-2016.html}{http://www.emis.de/journals/SIGMA/moduli-spaces-2016.html}}}

\Author{Chaiho RIM}

\AuthorNameForHeading{Chaiho~Rim}

\Address{Department of Physics, Sogang University, Seoul 121-742, Korea}
\Email{\href{mailto:rimpine@sogang.ac.kr}{rimpine@sogang.ac.kr}}

\ArticleDates{Received December 02, 2016, in f\/inal form February 27, 2017; Published online March 03, 2017}

\Abstract{We present recent developments of irregular conformal conformal states. Irregular vertex operators and their adjoint in a new formalism
are used to def\/ine the irregular conformal states and their inner product instead of using the colliding limit procedure. Free f\/ield formalism can be augmented by screening operators which provide more degrees of freedom. The inner product is conveniently given as the partition function of an irregular matrix model. (Deformed) spectral curve is the loop equation of the matrix model at Nekrasov--Shatashivili limit. We present the details of analytic structure of the spectral curve for Virasoso symmetry and its extensions, $W$-symmetry and super-symmetry.}


\Classification{11E04; 14H45; 15B52}

\renewcommand{\thefootnote}{\arabic{footnote}}
\setcounter{footnote}{0}

\section{Introduction}

Virasoro symmetry is typically represented by Virasoro module which consists of primary state and its descendants. An irregular conformal state is a dif\/ferent representation of the Virasoro symmetry and is a simultaneous eigenstate of a certain positive modes of conformal generators. Therefore, the irregular state is not a primary or descendant state and is rather a coherent state, combination of primary and its descendants. For example, the simplest irregular state is called Whittaker state~\cite{Whittaker}, the eigenstate of Virasoro $L_{+1}$ mode, and later generalized by Gaiotto~\cite{G_2009}. We use the def\/inition of the Virasoro irregular state of rank~$m$ as the simultaneous eigenstate of Virasoro generators~$L_k$ with $m \le k \le 2m$. This def\/inition can be similarly extended to $W$-symmetry; the simultaneous eigenstate of $W$-generators of spin~$s$; $W^{ (s+1)}_k$ with $s m \le k \le (s+1)m$. One can also construct the irregular states with the super-symmetry.

The irregular state is constructed as the superposition of a primary state and its descendants in \cite{BMT_2011,KMST_2013, MMM_2009}. However, there appear some ambiguities for the rank greater than~1. The superposition of states is not completely f\/ixed by the def\/ining relations, simultaneous eigenstate of a certain positive Virasoro generators. One of the main reason for this insuf\/f\/iciency is that other positive generators of the rank~$m$, $W^{ (s+1)}_k $ with $0 \le k <s m $ are non-commuting each other and require consistency condition when applied to the irregular state. It is also noted in \cite{CRZ_2014,CRZ_2015} that the consistency allows more degrees of freedom other than the simultaneous eigenvalues. Therefore, a~non-trivial systematic tool is needed to f\/ix the consistency conditions.

The irregular state is introduced in physics community in relation with AGT conjecture \cite{AGT_2010}. AGT connects Nekrasov partition function
of $N=2$ super Yang--Mills theory in 4 dimensions~\cite{Nekrasov} with the Liouville regular conformal block in~2 dimensions. At IR limit, one can have a~dif\/ferent class of conformal states, so called Argyres--Douglas theory of $N=2$ super Yang--Mills theory~\cite{AD}, which is not the deformation of UV conformal theory. It was pointed out in \cite{G_2009} that the Argyres--Douglas theory can be described in terms of the irregular state. In fact, a~simple way to obtain the irregular state is suggested in \cite{EM_2009,GT_2012}, which uses `colliding limit' of the Liouville conformal block or equivalently from Penner-type matrix model~\cite{DV_2009,IO_2010}.

During the last few years, even though many of the technical developments are achieved for the evaluation of irregular conformal block (ICB) in a signif\/icant way, the clear understanding of the irregular conformal block is still missing. The role of the colliding limit is not well understood. In this paper, instead of directly using the `colliding limit', we try a new approach to f\/ind ICB. We hope this new alternative approach provides a more intuitive way to understand the structure of ICB and Argyres--Douglas theory.

To do this, we adopt the irregular state of bosonic free f\/ields, which is directly obtained from the def\/inition of the irregular vertex operator
(IVO) \cite{GF_2014,NS_2010, PR_IV_2016}. The next step is to f\/ind the adjoint of the state, a necessary ingredient to def\/ine its inner product.
However, any explicit form of the adjoint of the irregular state has never been def\/ined in the literature. We present here the def\/inition of the adjoint, which uses the conformal transformation $z \to 1/z$ as used in the usual def\/inition of the adjoint of the regular primary state.
There needs, however, a proper rescaling of the state for the irregular one, which is the main dif\/ference from the usual def\/inition of the regular one. After this setup, the inner product is easily evaluated, and turns out to be f\/ixed by coherent coordinates only. This shows that the free f\/ield formalism is too restrictive to describe the Argyres--Douglas theory, since more degrees of freedom are needed. In fact, the necessary degrees of freedom are provided by the screening operators. As a result, the inner product is ef\/fectively given in terms of the partition function
of a beta-deformed Penner-type matrix model with the additional potential of the polynomial and inverse polynomial type, which is called the irregular matrix model in short. According to this set-up, the necessary building blocks of ICB are the irregular states, their adjoints and screening operators. It is demonstrated that all the previous results appeared for ICB are reproduced in this approach without using the colliding limit.

This paper is organized as follows. Section~\ref{sec:2} is the new set up of the formalism of ICB. Using the one bosonic free f\/ield formalism, the main idea is presented in a formal way. The irregular vertex operator of the free bosonic f\/ield is employed and its adjoint are newly def\/ined. The inner product between the irregular state and its adjoint is explicitly evaluated. This idea can be easily extended to many f\/ield formalism including fermionic f\/ields. After the set up, we include the screening operators to provide the inner product with more degrees of freedom. It is demonstrated that the ef\/fect of including screening operator in the inner product is conveniently represented as the partition function of an irregular matrix model. The result is equivalent to the colliding limit of the Penner-type matrix model which describes the regular conformal block.

Some explicit examples are given in Section~\ref{sec:3} and~\ref{sec:4}. The Virasoro symmetry is considered in Section~\ref{sec:3}. The inner product of the irregular conformal states coincides with the colliding limit of the Liouville regular conformal block. Therefore, one can use the technique developed in the random matrix formalism, namely, the loop equation which encodes the conformal symmetry of the irregular matrix model. For simplicity, we are considering the Nekrasov--Shatashivili (NS) limit~\cite{NS_2009}, where the loop equation is truncated to the one-point resolvent only. The resulting loop equation is viewed as the (deformed) spectral curve. The analytic structure of the spectral curve provides all the information of the system, including the partition function known as tau function. We discuss the analytic properties of the spectral curve, nature of the cut structure/pole structure, and the f\/low equation which plays the important role of f\/inding the partition function. One may extend the same method to $W$- and super-symmetry, which is shortly described in Section~\ref{sec:4}. Section~\ref{sec:5} is the conclusion.

We skip many of technical details since the purpose of this paper is to describe the structure of the irregular conformal block in a more intuitive way. Section~\ref{sec:2} and~\ref{sec:3} are enough to read for understanding the mainstream of ICB. Dedicated readers may found more
technical details in~\cite{CR_2015, CRZ_2015} for $W$-symmetry, and \cite{PR_SUSY_spectral_2016, PR_SUSY_2016} and references therein for supersymmetry.

\section[Irregular vertex operator, irregular state and its inner product]{Irregular vertex operator, irregular state\\ and its inner product}\label{sec:2}

\subsection{free f\/ield formalism}

General formalism of free f\/ield representation of the irregular vertex operator (IVO) $I^{(m)}(z)$ is considered in~\cite{GF_2014,NS_2010,PR_IV_2016} using bosonic f\/ield~$\phi$. In this section, we consider IVO of rank~$m$ using one f\/ield:
\begin{gather}
I^{(m)}(w) = e^{ 2 \Phi^{(m)}/\hbar}(w), \qquad \Phi^{(m)}(w) = \sum_{k=0}^m \frac{c_k}{k!} \frac{\partial^k \phi (w)}{\partial w^k},
\label{IVO-free-boson}
\end{gather}
where $\hbar$ is a convenient parameter
and the bosonic f\/ield has the holomorphic normalization
\begin{gather*}
\langle \phi(z) \phi (w) \rangle = -\frac12 \log(z-w), \qquad \big\langle e^{\alpha_1 \phi(z)} e^{\alpha_2 \phi} (w) \big\rangle = (z-w)^{-2 \alpha_1 \alpha_2}.
\end{gather*}
The energy momentum tensor and its moments are given as
\begin{gather}
T(z) = - (\partial \phi)^2 , \qquad L_k = \oint \frac{dz}{2 \pi} z^{1+k} T(z)\label{T(z)-free}
\end{gather}
and IVO satisf\/ies the operator product
\begin{gather}
\oint_{w=0} \frac{dz}{2 \pi} z^{1+k} T(z) I^{(m)} (w) =\Lambda_k ~I^{(m)},\label{T(z)-IVO}
\end{gather}
where $ \hbar^2 \Lambda_k = -\sum_{t+r=k} c_t c_r $. This shows that $ \Lambda_k=0 $ when $ k > 2m$.

Irregular state $| I^{(m)} \rangle$ is obtained if IVO is applied on the vacuum $|0 \rangle $:
\begin{gather}
| I^{(m)} \rangle =\lim_{z \to 0} I^{(m)}(z) |0 \rangle = I^{(m)}|0 \rangle. \label{IVO-state}
\end{gather}
The condition \eqref{T(z)-IVO} shows that the irregular state is the simultaneous eigenstate of positive generators
\begin{gather*}
\big[ L_k, I^{(m)} \big] =\Lambda_k ~I^{(m)} .
\end{gather*}
Therefore, $| I^{(m)} \rangle$ in \eqref{IVO-state} can be the simultaneous eigenstate of positive mode Virasoro gene\-ra\-tors~$L_a $ with~${m \le a \le 2m}$ since $L_{a \ge 0}$ satisf\/ies the commutation relation $[ L_a , L_b] =(a-b) L_{a+b}$.

The f\/ield derivative terms in IVO represent the descendant contributions. And the coef\/f\/i\-cients $c_k$ correspond to the coherent coordinates of the Heisenberg positive modes~$a_k$ with $1 \le k \le m$. This can be seen if one expands the bosonic f\/ield in terms of the Heisenberg modes $\partial \phi (z) = - \sum_k {a_k }/{z^{k+1}}$ and
\begin{gather*}
\oint_{w=0} \frac{dz}{2 \pi} z^{1+k} \partial \phi (z) I^{(m)} (w) =c_k I^{(m)}
\end{gather*}
or $[ a_k, I^{(m)}] = c_k I^{(m)}$. Therefore, it is obvious that
\begin{gather*}
a_k | I^{(m)} \rangle = c_k | I^{(m)} \rangle, \qquad 1 \le k \le m.
\end{gather*}

To def\/ine an inner product we need an adjoint state. The corresponding adjoint operator is def\/ined at inf\/inity
\begin{gather*}
\langle I^{(n)}|= \lim_{\zeta \to 0} \langle 0 | \hat I^{(n)} ( \zeta),
\end{gather*}
where conformal-transformation from the inf\/inity to 0, $z \to 1/\zeta$ is used. The adjoint operator is equivalent to the same one~\eqref{IVO-free-boson} but is def\/ined at $ 1/\zeta$
\begin{gather*}
\hat I^{(n)} (\zeta) =R_0^{(n)} e^{2 \hat \Phi^{(n)}/\hbar } (1/\zeta), \qquad \hat\Phi^{(n)} = \sum_{\ell=0}^n \frac{\hat c_{\ell}}{\ell!}
\frac{\partial^ \ell \phi (1/\zeta) }{\partial \zeta ^\ell}.
\end{gather*}
$R_0^{(n)}$ is the scaling factor
\begin{gather}
R_0^{(n)} = e^{2\frac{\hat c_0}{\hbar^2} \sum\limits_{\ell=0}^n \frac{\hat c_{\ell}}{\ell!} \frac{\partial^ \ell \log (\zeta) }{\partial \zeta ^\ell}} =
(\zeta)^{2 \frac{\hat c_0^2}{\hbar^2} } \prod_{\ell=1}^n e^{2 \frac {\hat c_0 \hat c_\ell}{\hbar^2 \ell! } \frac{\partial^ \ell \log(\zeta) }{ \partial \zeta^\ell} }.\label{R_0-scaling}
\end{gather}

Using the state and its adjoint we can def\/ine the inner product $\langle I^{(n)}| I^{(m)} \rangle $ which is given as two-point correlation
\begin{gather*}
\langle I^{(n)}| I^{(m)} \rangle_0 = \lim_{w, \zeta \to 0} \langle 0| \hat I^{(n)}(\zeta ) I^{(m)}(w) |0 \rangle_0.
\end{gather*}
We put the subscript 0 to emphasize that the inner product is def\/ined in terms of the free f\/ield representation. It is noted that the primary operator product has the non-vanishing expectation value if $\hat c_0 + c_0 =0$. This will be called the neutrality condition for the inner product.
The inner product need be symmetric
\begin{gather*}
\langle I^{(n)}| I^{(m)} \rangle_0=\langle I^{(m)}| I^{(n)} \rangle_0 .
\end{gather*}
Explicit evaluation shows the symmetric property
\begin{gather}
\langle I^{(n)}| I^{(m)} \rangle_0 = \lim_{w, \zeta \to 0} \exp\left (-2 \sum_{k,\ell \ge 1} \frac{\hat c_{\ell} c_k}{\ell! k!}
 \partial_\zeta^\ell \partial_w^k \log(1-w \zeta) \right)=e^{\zeta_{n,m}/\hbar^2},\nonumber\\
{\zeta_{n,m}}= \sum_{ \ell \ge 1} 2{\hat c_{\ell} c_\ell}/{\ell},\label{inner-free}
\end{gather}
where the summation in $ {\zeta_{n,m}}$ holds until $\ell \le {{\rm min} (m,n)}$. The scaling factor $R_0^{(n)}$ compensates the inf\/inite contribution at $\zeta =0$ so that the inner product is f\/inite. In addition, $\langle I ^{(0)}| I ^{(m)} \rangle_0 =\langle I^{(n)}| I ^{(0)} \rangle_0 =1$.
This is consistent with the fact that $| I ^{(m)} \rangle$ consists of primary state $| I ^{(0)} \rangle$ and its descendants where $\langle I ^{(0)}| I ^{(0)} \rangle_0$ is normalized to~1.

The inner product is the function of coherent coordinates $c_k$ and $\hat c_\ell$. We may use $c_k$ and $\hat c_\ell$ as the coordinates to represent the positive Virasoro mode acting on the state. For the case with $0 \le a \le m-1$, we have
\begin{gather}
L_a | I ^{(m)} \rangle = v_a | I ^{(m)} \rangle, \qquad v_a=\sum_k k c_{a+k} \frac{\partial}{\partial c_k}\label{eq:va}
\end{gather}
and $v_a$ satisf\/ies the commutation relation
\begin{gather}
[ v_a , v_b] =(b-a) v_{a+b}, \label{eq:va-comm}
\end{gather}
which corresponds to right action of the Virasoro commutation relation. One can check the commutation relation holds for the inner-product
\begin{gather*}
[ v_a , v_b] \langle I^{(n)}| I ^{(m)} \rangle =(b-a) v_{a+b} \langle I^{(n)}| I ^{(m)} \rangle \qquad \text{or}\qquad
[ v_a , v_b] (\zeta_{n,m}) =(b-a) v_{a+b}(\zeta_{n,m}).
\end{gather*}

Likewise, the adjoint state introduces the Virasoro representation in $\hat c_\ell$ space
\begin{gather}
\langle I ^{(n)}| L_a = \hat v_a~\langle I^{(n)}|, \qquad \hat v_a= \sum_\ell \ell \hat c_{a+\ell} \frac{\partial} {\partial \hat c_\ell},
\label{eq:hat-va}
\end{gather}
and $\hat v_a$ satisf\/ies the commutation relation
\begin{gather*}
 [ \hat v_a , \hat v_b] = (b-a) \hat v_{a+b}.
\end{gather*}
It is obvious that the commutation relation also holds for the inner-product
\begin{gather*}
[\hat v_a , \hat v_b] (\zeta_{n,m}) =(b-a) \hat v_{a+b}(\zeta_{n,m}).
\end{gather*}
Finally, it is noted that the two representations $\hat v_a$ and $v_a$ are commuting each other
\begin{gather*}
[v_a, \hat v_b]=0.
\end{gather*}
It is noted that the f\/irst-order dif\/ferential representation of the positive Virsoro generators allows the calculation much easier. On the other hand, if one wants to consider the full representation of the positive and negative Virsoro generators, then one inevitably needs to consider the second-order dif\/ferential representation.

\subsection{Screening operator and irregular matrix model}

Whittaker state, the eigenstate of $L_1$ and annihilated by $L_{k>1}$ was constructed explicitly in~\cite{Whittaker} as the superposition of a primary state and its descendants. In \cite{G_2009,MMM_2009}, the irregular state of rank~1 is also constructed. And later, the irregular state of arbitrary rank $m >1$ is suggested in \cite{KMST_2013} with unknown coef\/f\/icients
\begin{gather}
| {G_{2m}} \rangle = \sum_{\ell,Y, \ell_p} \Lambda^{\ell/m} \left\{ \prod_{i=1}^{m-1} a_i ^{\ell_{2m-i} } b_i^{\ell_i} \right\}\nonumber\\
\hphantom{| {G_{2m}} \rangle =}{}\times
\mu^{\ell_m} Q_{\Delta} ^{-1} \big( 1^{\ell_1}2^{\ell_2} \cdots (2m-1) ^{\ell_{2m-1}} (2m) ^{\ell_{2m}}; Y \big) L_{-Y} |\Delta \rangle,
\label{G_2m}
\end{gather}
where $ |\Delta \rangle $ is the primary state with conformal dimension $\Delta$. $L_{-Y}=L_Y^+ $ represents the product of lowering operators and $L_Y = L_1^{\ell_1} L_2^{\ell_2} \cdots L_s^{\ell_s}$. The summation $\ell$ and $Y$ run from 0 to $\infty$, maintaining $\ell =|Y|=\sum_p p \ell_p $. $Q_{\Delta} (Y;Y')$ is the Shapovalov form, $Q_{\Delta} (Y;Y')= \langle \Delta | L_{Y'} L_{-Y}|\Delta \rangle$. The coef\/f\/icients $a_i$ and $\mu$ are closely related with eigenvalues; $\Lambda_{2m} = \Lambda^m$, $\Lambda_k = \Lambda^{k /m} a_{2m-k} $ for $m < k < 2m$ and $ \Lambda_m =\Lambda \mu $. However, $b_i$'s are not f\/ixed by the eigenvalues but still gives the contribution to the expectation values of the lower positive generators. The coef\/f\/icients are to be f\/ixed with the consistency conditions given in~\eqref{eq:va-comm}. This suggests that free f\/ield IVO
cannot be a unique solution. In addition, considering AGT which relates CFT with $N=2$ super gauge theory, we need more degrees of freedom to connect the irregular states with Argyres--Douglas theory.

More degrees of freedom can be added when homogeneous solution $O$ satisfying $[ L_k, O] =0$ for $k \ge 0$ is considered. The solution $O$ is called screening operator and is given as a contour integral of primary operator $\psi(z)$ with holomorphic dimension~1
\begin{gather*}
O= \oint dz \psi (z) .
\end{gather*}
Then, the screening operator allows a general solution which will be the free f\/ield solution \eqref{IVO-free-boson} with any number of screening operators multiplied.

Liouville f\/ield theory has the scaling dimension 1 operator in the presence of background charge $Q=b + 1/b$. The vertex operator $V_\alpha (z) =e^{2 \alpha \phi}(z)$ has the scaling dimension $\Delta_\alpha = \alpha (Q-\alpha)$ since the background charge modif\/ies the energy-momentum tensor from the free f\/ield case \eqref{T(z)-free}
\begin{gather*}
T(z) = - (\partial \phi)^2 + Q \partial^2 \phi .
\end{gather*}
The background charge makes the conformal dimension of $V_b (z)$ be 1 and changes the eigen\-va\-lue~$\Lambda_k$ of IVO appeared in~\eqref{IVO-free-boson}
\begin{gather}
\hbar^2 \Lambda_k =\epsilon (k+1) c_k - \sum_{r+s=k} c_r c_s,\label{Lambda_k}
\end{gather}
where we use the notation $\epsilon= \hbar Q$.

Due to the insertion of the screening operators, the inner product becomes more complicated
\begin{gather*}
\langle I ^{(0)}| I ^{(m)} \rangle_N= \langle I ^{(0)}| O^N | I ^{(m)} \rangle_0 = Z^{(0,m)}_N [c_0; \mathbf {c}],
\end{gather*}
where $N$ denotes that there are $N$ number of screening operators are inserted. $Z^{(0,m)}_N$ is the partition function of irregular matrix model
\cite{EM_2009,NR_2012} and depends on the coherent coordinates $\mathbf {c} = \{c_1, \dots, c_n\}$:
\begin{gather}
Z^{(0,m)}_N [ c_0; \mathbf {c}] =\left[ \int \prod_{I=1}^N d\lambda_I \right] \prod_{I<J}^N (\lambda_I- \lambda_J)^{2\beta } e^{ \frac{\sqrt{\beta}}g V^{(0,m)} ( \{\lambda_I \} )},\nonumber\\
V^{(0,m)}(\{ \lambda_I\} ) = \sum_{I=1}^N \left\{ c_0 \log (\lambda_I ) - \sum_{k=1}^m \frac{c_k}{k \lambda_I^k} \right \}. \label{Z_0m}
\end{gather}
Here we redef\/ine $\beta = -b^2$ or $b=i \sqrt{\beta}$ and $g = i \hbar/2$ for later convenience.

When no screening operator is included, the inner product $\langle I^{(0)} | I^{(m)} \rangle_N$ reduces to $\langle I^{(0)} | I^{(m)} \rangle_0$ $ =1$ as in free f\/ield theory formalism in~\eqref{inner-free}. However, it should be emphasized that $\langle I^{(0)} | I ^{(m)} \rangle_N$
is not normalized as~1 due to the screening operator ef\/fect and the parameter dependence of the inner product is of the primary concern.

The inner product $\langle I ^{(0)}| I ^{(m)} \rangle_N$ should be symmetric
\begin{gather*}
\langle I ^{(0)}| I ^{(m)} \rangle_N =\langle I ^{(m)}| I ^{(0)} \rangle_N,
\end{gather*}
where the proper scaling factor in the def\/inition of the adjoint def\/inition \eqref{R_0-scaling} is modif\/ied due to the background charge
\begin{gather}
R_\epsilon^{(m)} = e^{2 \frac{(\hat c_0-\epsilon )}{\hbar^2} \sum\limits_{\ell=0}^m \frac{\hat c_{\ell}}{\ell!} \frac{\partial^ \ell \log ( \zeta) }{\partial \zeta ^\ell} } =(\zeta)^{2 \frac{(\hat c_0 -\epsilon)\hat c_0 }{\hbar^2}} \prod_{\ell=1}^m e^{ 2 \frac { (\hat c_0 -\epsilon) \hat c_\ell}{\hbar^2 \ell! } \frac{\partial^ \ell \log(\zeta) }{ \partial \zeta^\ell} }. \label{R_e-scaling}
\end{gather}
Note that the symmetric property of the inner-product is related with the conformal invariance. After the conformal transformation $\lambda_I \to 1/ \lambda_I$, the partition function changes to $Z^{(m,0)}_N [\hat c_0; \mathbf {c}]$ whose potential is now given as
\begin{gather*}
V^{(m,0)}(\{ \lambda_I\} ) = \sum_{I=1}^N \left\{ \hat c_0 \log (\lambda_I ) - \sum_{k=1}^m \frac{\hat c_k \lambda_I^k}{k}\right \},
\end{gather*}
where $\hat c_k =c_k$ and $\hat c_0$ is a quantity hidden at inf\/inity and is def\/ined through the neutrality condition
\begin{gather}
\hat c_0 +\hbar Nb +c_0 = \epsilon.\label{eq:neutrality}
\end{gather}
Conformal invariance of the partition function shows that
\begin{gather*}
Z^{(0,m)}_N [c_0; \mathbf {c}] = Z^{(m,0)}_N [\hat c_0; \hat {\bf c}]
\end{gather*}
and therefore, the symmetric property of the inner product is assured.

The inner product of two irregular states is similarly given as
\begin{gather*}
\langle I ^{(n)}| I ^{(m)} \rangle_N ={\cal Z}^{(n,m)}_N [ c_0;\mathbf {\hat c}; \mathbf {c}] = \lim_{w, z \to 0} \langle 0| I ^{(n)}(1/w)~ O^N I ^{(m)}(z) |0 \rangle = e^{\zeta_{n,m}/\hbar^2 } Z^{(n,m)}_N[ c_0;\mathbf {\hat c}; \mathbf {c}],
\end{gather*}
where ${\zeta_{n,m}}$ is the free f\/ield contribution and $Z^{(n,m)}_N$ is the screening f\/ield ef\/fect~\cite{CR_2013}
\begin{gather}
Z^{(n,m)}_N[c_0; \mathbf {\hat c}, \mathbf {c}] = \left[ \int \prod_{I=1}^{N } d\lambda_I \right] \prod_{I<J}^{N} (\lambda_I- \lambda_J)^{2\beta }
e^{ \frac{\sqrt{\beta}}g V^{(n,m)} ( \{\lambda_I \} )}, \label{Z_nm}\\
V^{(n,m)}(\{ \lambda_I\} ) = \sum_{I=1}^N \left\{ c_0 \log (\lambda_I ) - \sum_{k=1}^m \frac{c_k}{\lambda_I^k}-\sum_{\ell=1}^n \frac{\hat c_\ell}{\ell} \lambda^\ell \right \}. \label{potential-nm}
\end{gather}
The conformal invariance
\begin{gather*}
{ Z}^{(n,m)}_N[ c_0; \mathbf {\hat c}, \mathbf {c}] ={ Z}^{(m,n)}_N [ \hat c_0; \mathbf {c}, \mathbf {\hat c}]
\end{gather*}
is obvious with the neutrality condition \eqref{eq:neutrality} and so is the symmetric property of the inner product.

One may also def\/ine irregular conformal block ${\cal F}_{N_1, N_2}^{(n,m)} [ c_0; \mathbf {\hat c}, \mathbf {c}]$ in terms of inner product~\cite{CRZ_2015}
\begin{gather*}
\langle I^{(n)} | I^{(m)} \rangle_N =\sum_\Delta {\cal F}_{N_1, N_2}^{(n,m)} [ c_0; \mathbf {\hat c}, \mathbf {c}] ~\langle I^{(n)} | \Delta \rangle_{N_1} \langle \Delta | I^{(m)} \rangle_{N_2},
\end{gather*}
where $N=N_1+N_2$ and $ | \Delta \rangle$ is an intermediate primary state of conformal dimension $\hbar^2 \Delta =(\hat c_0 + N_1 \epsilon) (c_0 + N_2 \epsilon)$. In this def\/inition, the intermediate states are to be inserted in the def\/inition of the inner product $\langle I^{(n)} | I^{(m)} \rangle_N$. The irregular conformal block is given in terms of the partition function as
\begin{gather*}
 Z^{(n,m)}_N [ c_0; \mathbf {\hat c}, \mathbf {c}] =e^{-\zeta_{n,m}/\hbar^2 }\sum_{N_1+ N_2 =N} {\cal F}_{N_1, N_2}^{(n,m)} [ c_0; \mathbf {\hat c}, \mathbf {c}] Z^{(n,0)}_{N_1}[ \hat c_0; \mathbf {\hat c} ] Z^{(0,m)}_{N_2}[c_0; \mathbf {c}] . 
\end{gather*}
One practical way of evaluating ICB is to use the perturbative approach \cite{CRZ_2014}, where the potential of partition function $ Z^{(n,m)}_N [ c_0; \mathbf {\hat c}, \mathbf {c}]$ is splitted so that the reference one is given as $Z^{(n,0)}_{N_1}[ \hat c_0; \mathbf {\hat c} ] \times
Z^{(0,m)}_{N_2}[c_0; \mathbf {c}]$ and the rest as the perturbative one. The explicit calculation demonstrates that the parameter~$b_k$ appeared in~\eqref{G_2m} is identif\/ied as $\Lambda^{k/m} b_k = \Lambda_k + v_k \big({-}\hbar^2 \log Z_N^{(0,m)} \big)$ with $\Lambda^m= \Lambda_{2m}$.

The irregular matrix model~\eqref{Z_nm} can also be obtained using the colliding limit~\cite{CR_2013} of he regular Liouville conformal block, holomorphic correlation of vertex operators $V_{\alpha} =e^{2 \alpha \phi}$ together with the screening operators. The regular conformal block
with $(n+m+2) $ primary operators
\begin{gather*}
{\cal G }^{(n+m)}_N = \left\langle \prod_{A=1}^{n+m+2} V_{\alpha_A} (z_A) \left( \oint dz ~e^{b \phi}(z) \right)^N \right\rangle
\end{gather*}
satisf\/ies the neutrality condition $ \sum\limits_{A=1}^{n+m+2} \alpha_A + bN = Q$. Colliding limit is to put $m+1$ operators at the origin with $c_k $ $( 1 \le k \le m) $ f\/inite
\begin{gather*}
c_k =\sum_{A=1}^m \hbar \alpha_A z_A^k,
\end{gather*}
where $z_A$ is the coordinates of the operators approaching to 0. We rescaled $\alpha_A$ with $\hbar$ so that $c_k$ is identif\/ied with the coef\/f\/icients in the potential in~\eqref{potential-nm}. At inf\/inity, $n+1$ primary operators is put to inf\/inity with $\hat c_\ell$ $( 1 \le \ell \le n) $ f\/inite
\begin{gather*}
\hat c_\ell= \sum_{A=1}^n \hbar \alpha_A \zeta_A ^{\ell}
\end{gather*}
with $ z_A=1/\zeta_A \to\infty$. After the proper scaling $R_\epsilon ^{(n)} $ introduced in \eqref{R_e-scaling}, ${\cal G }^{(n+m)}_N $ reduces to
the inner product
\begin{gather*}
\langle I^{(n)}| I^{(n)} \rangle_N = e^{\zeta_{n,m}/\hbar^2} {\cal Z}^{(n, m)}_N [ c_0; \mathbf {\hat c}, \mathbf {c}] .
\end{gather*}
It is to be noted that the free f\/ield contribution $e^{\zeta_{n,m}}$ is called $U(1)$ ef\/fect which appears from the factor
\begin{gather*}
\lim_{\zeta_A, z_B \to 0} \prod_{A,B} (1 -\zeta_A z_B)^{- 2 \alpha_A \alpha_B} =e^{\zeta_{n,m}/\hbar^2}
\end{gather*}
at the colliding limit $ (z_B, \zeta_A) \to 0$.

\section{Properties of inner product}\label{sec:3}

\subsection{(Deformed) spectral curve}

The inner product \eqref{Z_0m} or \eqref{Z_nm} requires a systematic evaluation of the ef\/fect of the screening operators. We are using the salient feature of the partition function, the conformal invariance. Under the holomorphic transform $\lambda_I \to f(\lambda_I) = \lambda_I + \delta/(z- \lambda_I)$, the conformal invariance of the partition function is encoded in the loop equation
\begin{gather}
4 W(z)^2 +4 V'(z) W(z) + 2 \epsilon W'(z) - \hbar^2 W(z,z) =f(z) .\label{eq:loop}
\end{gather}
Here $ W(z)$ is the one-point resolvent
\begin{gather}
 W(z) = g \sqrt{\beta} \left\langle\sum_I \frac{1} {z - \lambda_I} \right\rangle, \label{one-point-resolvent}
\end{gather}
where the bracket $ \langle \cdots \rangle$ denotes the expectation value with respect to the irregular matrix model. Primed quantity denotes the derivative with respect to its argument. Two point resolvent~$W(z,z)$ is given as the connected part
\begin{gather*}
 W(z_1, z_2) = {\beta} \left\langle \sum_I \frac{1} {z_1 - \lambda_I} \sum_J \frac{1} {z_1 - \lambda_J} \right\rangle_{\rm conn}.
\end{gather*}
$ f(z) $ has the potential dependence
\begin{gather*}
f(z) = 4g \sqrt{\beta} \left\langle \sum_I \frac{V'(z)- V'(\lambda_I)}{z - \lambda_I} \right\rangle .
\end{gather*}

For simplicity, we will consider the classical/NS limit of the loop equation~\eqref{eq:loop}. Nekrasov--Shatashivili (NS) limit is obtained if $\hbar \to 0$ and $b \to \infty$ so that $\epsilon = \hbar Q =2 g \sqrt \beta$ is f\/inite~\cite{NS_2009}. On the other hand, classical limit is obtained in the Liouville f\/ield theory when and $b \to 0$ and $\epsilon = \hbar Q $ is f\/inite. However, Liouville f\/ield theory has the duality $b \to 1/b$
so that two limits, $b \to \infty$ and $b \to 0$ are equivalent. At the classical/NS limit, one can check that multi-point resolvent vanishes and
the loop equation is given in terms of one-point resolvent only
\begin{gather}
4 W(z)^2 +4 V'(z) W(z) + 2 \epsilon W'(z) =f(z) . \label{loop-classical}
\end{gather}
This loop equation becomes a more informative form if $ x = 2 W + V' $ is used
\begin{gather}
x ^2 + \epsilon x' + \xi_2(z) =0. \label{virasoro-spectral-curve}
\end{gather}
This Riccati equation is regarded as a~deformed (due to $\epsilon x'$ term) spectral curve.

Analytic structure of the spectral curve is specif\/ied by $ \xi_2 (z) = - V'^2 +\epsilon V'' - f $, contains details of the information of Virasoro symmetry. This fact is seen as follows: For the type of potential $V=V^{(n,m)}$, $f$ has the analytic structure
\begin{gather*}
 f = \sum_{a=-n}^{m-1} \frac{d_a}{z^{a+2}},
\end{gather*}
where the prime in the summation denotes that $a=-1$ is missing since $d_{-1}=0$. Therefore, $\xi_2 (z) $ has irregular poles and zeros on the Riemann surface
\begin{gather*}
 \xi_2(z) =\sum_{k=-2n}^{2m} \frac{\hbar^2 \Lambda_k}{z^{k+2}} -\sum_{a=-{n}}^{m-1} \frac{d_a}{z^{a+2}},
\end{gather*}
where $ \Lambda_k$ is the same one given in~\eqref{Lambda_k} with the def\/inition $c_{-k}= \hat c_k$. Due to the irregular singularity, the Riccati equation~\eqref{virasoro-spectral-curve} shows the dif\/ferent analytic structure from the regular ones.

Suppose we ignore $f(z)$ (putting $d_a=0$). Then, $\xi_2(z)$ is given in terms of $\Lambda_k$. The positive modes $L_k$ for $m \le k \le 2m$ apply on the state $|I ^{(m)}\rangle$, and the negative modes $L_k$ for $-2n \le k \le -n $ on the state $\langle I ^{(n)}|$, both resulting in the desired eigenvalues.

To understand the role of $d_a$, let us consider the case $d_a$ with $a \ge0$. One can check that from the def\/inition of $f$ one has
\begin{gather}
d_a= v_a (F^{(n;m)} ) \qquad {\rm for} \quad 0\le a \le m-1, \label{flow-da}
\end{gather}
where $F_N^{(n;m)} =-\hbar^2\log Z_N^{(n;m)} $ and $v_a $ is given in \eqref{eq:va}, Virasoro representation in $\{c_k\}$ space. For the case $d_{-a}$ with $a \ge 2$ one has $d_{-a} = - 2 \epsilon \hat c_a + \hat d_a $. We have $ \hat d_a $ with $ { 0 \le a \le n-1}$ in terms of Virasoro representation
\begin{gather}
\hat d_a = \hat v_a \big(F_N^{(n;m)} \big), \label{flow-hat-da}
\end{gather}
where $\hat v_a $is given in \eqref{eq:hat-va}, Virasoro representation in $\{\hat c_{\ell}\}$ space. Therefore, it is clear that $\xi_2(z) $ is identif\/ied with the expectation value of the energy momentum tensor (Virasoro current)
\begin{gather*}
 \xi_2 = \langle \hbar^2 T(z) \rangle =\sum_{k \in \mathbb{Z}} \frac{ \langle \hbar^2 L_k \rangle }{z^{k+2}} .
\end{gather*}
Here, expectation is given in terms of irregular matrix model and is also considered as the normalized expectation value
\begin{gather*}
\langle A \rangle = \frac{\langle I^{(n)} | A | I^{(m)} \rangle_N} {\langle I^{(n)} | I^{(m)} \rangle_N}.
\end{gather*}

It is also important to note that $\hbar^2 \Lambda_k$ and the free energy $F^{(n;m)} $ is regarded as f\/inite at the classical/NS limit~\cite{RZ_2015}. Moreover, since $ Z_N^{(n;m)} = \exp(- F^{(n;m)}/\hbar^2)$, the f\/low equations~\eqref{flow-da} and~\eqref{flow-hat-da} satisf\/ies consistency conditions
\begin{gather}
 v_a (d_b) - v_b (d_a) = (a-b ) d_{a +b }, \qquad \hat v_a ( \hat d_b) - \hat v_b ( \hat d_a) = (a-b ) \hat d_{a +b }, \nonumber\\
 v_a ( \hat d_b) - \hat v_b ( d_a) = 0 .\label{flow-consistency}
\end{gather}
The consistency conditions are useful tools to construct the partition function. The major step is to f\/ind the values $d_a$ and $\hat d_a$ directly from the analytic property of the spectral curve~\eqref{virasoro-spectral-curve}. This procedure is presented in the following subsection.

\subsection{Irregular spectral curve and polynomial equation}

When $\epsilon =0$, the spectral curve \eqref{virasoro-spectral-curve} becomes the large $N$ limit of random matrix models which can be solve as in the usual approach~\cite{C_2010, CEM_2011}. Therefore, one can solve the spectral curve keeping~$x $ is~$ O(1)$ so that~$\epsilon x'$ is regarded as sub-dominant. In this case, the dominant term is simply given as $ x = \pm \sqrt{- \xi_2} $ and solves the spectral curve~\eqref{virasoro-spectral-curve} as the perturbation series in powers of~$\epsilon$~\cite{CR_2013, NR_2012}. As a result, the solution has $(m+n)$ square-root branch cuts and provides the double covering of the Riemann surface~\cite{DV_2002, DV_2009}.

However, at the classical/NS limit, the derivative term $x'$ survives in the spectral curve \eqref{virasoro-spectral-curve} and can change the analytic structure of the solution space. This is related with the well-known fact that one can transform the Riccati equation into a second-order linear dif\/ferential equations. Putting $x = \epsilon (\log \Psi)'$ one has
\begin{gather*}
\left (\epsilon^2 \frac{\partial^2}{\partial z^2} + \xi_2 (z) \right) \Psi (z) =0.
\end{gather*}

To understand the meaning of the transformation of the spectral curve, let us consider an expectation value \cite{BMT_2011b, RZ}
\begin{gather}
P(z) \equiv \left\langle \prod_I ( z-\lambda_I ) \right\rangle = \sum_{A=0}^N P_A z^A = \prod_\alpha (z-z_\alpha), \label{eq:PA}
\end{gather}
which is a monic polynomial of degree $N$ ($P_N=1$). $z_\alpha$ are zeros of the polynomial, which are assumed distinct. One can check that $P(z)$ is related with the resolvent~$W(z)$ in~\eqref{one-point-resolvent} at the classical/NS limit. Using $\log \big(\prod_I ( z-\lambda_I )\big) \propto \sum _I \int^z \frac{dz'}{ z'-\lambda_I } $ one has
\begin{gather}
\log \left( \frac { P (z)}{ P (z_0)} \right) = \frac2 {\epsilon } \int_{z_0}^z dz' W(z' ), \label{eq:P-W}
\end{gather}
since the multi-point resolvent vanishes at the classical/NS limit \cite{BMT_2011b, MMM_2011}. Taking the derivative of~\eqref{eq:P-W}, we have
\begin{gather}
2 W(z) = \epsilon { P' (z)}/ { P (z)} = \epsilon \sum_{\alpha=1}^N \frac 1 {z-z_\alpha}.
\label{pole-position}
\end{gather}
Then, the monic polynomial $P(z)$ satisf\/ies the second-order dif\/ferential equation
\begin{gather}
{\epsilon}^2 P''(z) + 2{\epsilon} V'(z) P'(z) = f(z) P(z). \label{eq:P}
\end{gather}
The solution of the dif\/ferential equation shows that there are $N$ zeros and therefore, the resolvent $W(z)$ has simple poles. As a result, one can conclude that there will appear $N$-simple poles in the spectral curve rather than the branch cut present when $\epsilon=0$. The branch cut disappears and only simple poles are present.

In addition, one can f\/ind $d_a$ from the dif\/ferential equation~\eqref{eq:P} in terms of coherent coordinates, $\{ c_k\}$ and $\{\hat c_k\}$.
Then, the consistency condition \eqref{flow-consistency} is solved by noting that $v_a = \sum_k U^{-1} _{ak}\partial_k$ or $\partial_k =\sum_a U_{ka} v_a$, where $\partial_k = \partial/ \partial c_k$. $U$ is a $m \times m $ matrix for the rank~$m$ case and is simply given by the coherent coordinates. Then, a closed one form $ d ( F^{(n,m)} ) = \sum_k d c_k ~ \partial_k ( F^{(n,m)}) $ satisf\/ies the f\/low equations $ d ( F^{(n,m)} ) = \sum_{k,a} d c_k ~ U_{ka} ~d_a$. According to the spectral curve~\eqref{loop-classical}, $d_a $ is the residue of $z^{1+a} \big(4 W^2(z) + 4 V'(z) W(z) +2 \epsilon W'(z)\big )$ so that
\begin{gather*}
d_a = - 4 \sum_{\ell \ge 1} \frac {c_{a+\ell} }{\ell !}~ \partial^\ell W(0) = 2 \epsilon \sum_{\ell \ge 1} c_{a+\ell}\left(\sum_{\alpha=1}^N \frac 1{z_\alpha^\ell} \right),
\end{gather*}
where the solution $W(0)$ is the function of coherent coordinates. Using the zeros of $P(z)$, one has the f\/low equation
\begin{gather}
d \big( F_N^{(n,m)} \big) = \sum_{k, a} d c_k U_{ka} d_a = 2 \epsilon \sum_{k, a} d c_k U_{ka}\sum_{\ell \ge 1} c_{a+\ell}\left(
\sum_{\alpha=1}^N \frac 1{z_\alpha^\ell} \right). \label{eq:dF}
\end{gather}
The same form of the f\/low equation holds for the hat coordinates
\begin{gather*}
\hat d \big( F_N^{(n,m)} \big) = \sum_{k, a} d \hat c_k \hat U_{ka} \hat d_a = 2 \epsilon \sum_{k, a} d \hat c_k \hat U_{ka}\sum_{\ell \ge 1} \hat c_{a+\ell} \left( \sum_{\alpha=1}^N \frac 1{\hat z_\alpha^\ell}\right).
\end{gather*}
where $\hat U $ and $\hat z_\alpha$ are the adjoint expression which is obtained from the $ U $ and $ z_\alpha$ by putting \mbox{$c_a \leftrightarrow \hat c_a$}. (The relation $\hat d_{a\ge 2} $ can be checked from the residue $z^{1-a} \big(4 W^2(z) + 4 V'(z) W(z) +2 \epsilon W'(z) \big)$).

It is to be noted that the polynomial $P(z)$ is closely related with the degenerate primary operator expectation value. Let us consider $\langle I^{(n)} |\,V_{+}(z)\,|\, I^{(m)} \rangle$, where $ V_+ (z) \equiv V_{ -1/(2b)} $ has the conformal dimension $ \Delta_{+} = -\frac{1}{2}-\frac{3}{4b^2} $ and has the null vector at level~2. Let us def\/ine $\Psi^{(n,m)}_+(z) $, normalized expectation value
\begin{gather*}
 \Psi^{(n,m)}_+ (z) = \frac{ \langle I^{(n)} |\,V_{+}(z)\,|\, I^{(m)} \rangle_{N_+} } { \langle I^{(n)} | I^{(m)} \rangle_{N }},
\end{gather*}
where $N _+ - N = 1/(2b^2)$ is assumed for the neutrality condition to hold. This requires that the number of screening operators used to evaluate the partition function is dif\/ferent from that used to evaluate the expectation value. At NS limit, however, this unpleasant feature disappears
since $N _+ =N$. Then it is easy to f\/ind
\begin{gather*}
 \Psi^{(n,m)}_+ (z) =P(z) e^{V^{(n;m)}(z) / \epsilon} = \exp \left( \frac{1}{\epsilon} \int^{z} x(z') dz' \right),
\end{gather*}
where $P(z)$ is the solution of \eqref{eq:P}. This show that the deformed spectral equation~\eqref{virasoro-spectral-curve} or~\eqref{eq:P}
leads to the second-order dif\/ferential equation of $\Psi^{(n,m)}_+ (z) $
\begin{gather*}
\left ( \epsilon^2 \frac{\partial^2}{\partial z^2} + \xi_2 (z) \right) \Psi^{(n,m)}_+ (z) =0,
\end{gather*}
which is exactly the second-order dif\/ferential equation obtained from the Riccati equation.

One may also use the null-vector appearing at level 2
\begin{gather*}
\chi_{+}(z)=\left[\widehat{L}_{-2}(z) - \frac{3}{2(2\Delta_{+} +1)} \widehat{L}_{-1}^{2}(z)\right]V_{+}(z).
\end{gather*}
The null vector vanishes when inner product is evaluated with any state. Therefore, one has the null constraint $\langle I^{(n)} |
\,\chi_{+}(z)\,|\, I^{(m)} \rangle =0$ which is represented in terms of the second-order dif\/ferential equation with respect to~$z$. It is shown in \cite{LLNZ_201309, PP_201407} that the equation is the Mathieu equation for the Whittaker case. If $m=n$, we have Generalized Mathieu equation on a circle with $z={\rm e}^{2ix}$ when the wave function $\Psi^{(n,m)}_+ (z) $ is properly rescaled~\cite{RZ}.

As a side remark, Liouville f\/ield theory has an equivalent screening operator
\begin{gather*}
\tilde O= \oint dz ~e^{ \phi/b}(z)
\end{gather*}
due to the dual symmetry under $b \to 1/b$. Therefore, one can equivalently uses either $O$ or $\tilde O$. We can f\/ind the ef\/fect of inserting an new screening operator~$\tilde O$ by evaluating the expectation value with respect to the partition function
\begin{gather*}
\tilde \Psi^{(n,m)} (z) = \frac{ \langle I^{(n)} |\, e^{ \phi/b}(z) \,|\, I^{(m)} \rangle_{\tilde N} } { \langle I^{(n)} | I^{(m)} \rangle_{N }},
\end{gather*}
where $\tilde N =N + 1/b^2$ and at NS limit, one has $\tilde N =N$. Noting that $\tilde \Psi^{(n,m)} (z)= 1/(\Psi^{(n,m)}_+ (z) )^2$, one has
\begin{gather*}
\langle \tilde O \rangle = \oint \frac{dz }{P^2(z)} e^{-\frac2 \epsilon V^{(n;m)}(z) } = \sum_k \left. \frac {2\pi i} {(dP/dz)^2 } \frac {d e^{-\frac2 \epsilon V^{(n;m)} }} {d z} \right|_{z= z_k},
\end{gather*}
where $z_k$ are the zeros of $P(z)$. One may also f\/ind that the expectation value of higher level degenerate operators is given as the higher powers of $\Psi^{(n,m)}_+ (z)$'s. However, simultaneous use of the two screening operators (or inf\/inite number of $\tilde O$'s) will break the conformal symmetry as seen in the sine-Gordon case, which needs further investigation.

\subsection{Filling fraction and branch cuts}

The deformed spectral curve is written as the second-order dif\/ferential equation~\eqref{eq:P} of a~monic polynomial $P(z)$. For the type of potential $V^{(0,m)}$ one may multiply \eqref{eq:P} by $z^{m+1}$ to f\/ind $N+m-1$ order of polynomial equation which provides $N+m$ independent relations. Noting the number of unkowns are $N+m$ ($N$ number of pole positions and $m$ number of $d_a$'s), the polynomial equations determine the unknown completely. For the case $V^{(n,0)}$, the same conclusion arises if one uses the dual potential $ V^{(0,n)}$ after the conformal transformation. For~$V^{(n,m)}$, we have $N+m+n-1$ order of polynomial equation and end up with $N+m+n$ independent relations. This is consistent with the fact that there are $N+m+n$ unknowns ($N$~number of pole positions and $m$ number of $d_{a\ge 0}$ and $n$ number of $d_{a< 0}$).
Therefore, the dif\/ferential equation\eqref{eq:P} completely f\/ixes~$d_{a\ge 0}$ and~$\hat d_{a \ge 0}$.

Once the solution is known, the partition function can be found through the f\/low equation~\eqref{eq:dF}. On the other hand, the branch cut structure is absent in the spectral curve. This raises a~question. In the regular case, the integration contour of the partition
function \eqref{Z_nm} is def\/ined so that the integration contour includes the branch cut. The partition function is f\/ixed according to the
distribution of integration contours around the branch cuts. On the other hand, how can one def\/ine the contour integral of the partition function
if the branch cut disappears?

The hint lies on the degrees of freedom in the solution space. To see this, let us consider the case $n=0$, $m=2$ and $N=1$. We have 3 unknowns, $P_0$, $d_0$ and $d_1$. $d_0$ is trivially given: $d_0 = 2 \epsilon c_0$. However, $d_1 = 2 \epsilon c_1 -d_0 P_0$ is f\/ixed by a quadratic equation
\begin{gather*}
d_1^2 -2 \epsilon c_1 d_1 + 2 \epsilon c_2 d_0 =0.
\end{gather*}
Therefore, $d_1$ has two solutions: $d_1^\pm = \epsilon c_1 \big( 1 \pm \sqrt{1-\eta} \big) $ where $\eta= 4c_2c_0/c_1^2$. Each solution corresponds to $P_0^- \sim c_1/c_0$ and $P_0^+ \sim c_2/c_1$ which shows that the root lies near one of two stationary points of the potential. In conclusion, zeros of the polynomial (or the poles of the resolvent) may distribute dif\/ferently around the dif\/ferent stationary points of the potential.
Accordingly, the dif\/ferent solution results in the dif\/ferent partition function.

As $N$ increases, the solution space of \eqref{eq:P} rapidly becomes very complicated. Suppose we consider the potential $V^{(0,m)}$. One can easily convince that the coef\/f\/icient $P_{N-1} $ in \eqref{eq:PA} has $N+1$ solutions. (For the case $m$ with $N$ zeros, one has $ \frac{(N+m-1)!}{N!(m-1)!}$ solutions. One may view this solutions as the zero distribution with $N=\sum\limits_{i=1}^m N_i$. $N_i$ is the number
of zeros around each stationary point of the potential.) Therefore, we need more ef\/f\/icient way to f\/ind $d_a$'a and $\hat d_a$'s.

Alternative approach is to use the f\/illing fraction from the beginning. One can f\/ind the number $N_a$ of inserted screening operators using the one point resolvent~$W(z)$ in \eqref{one-point-resolvent} or~\eqref{pole-position}
\begin{gather*}
{\epsilon} N_a = \oint_{{\cal A}_a} \frac{dz}{2 \pi i}~ 2W(z)
\end{gather*}
if the integration contour ${\cal A}_a$ locates around the saddle point of the potential. Therefore, the integration contours in the partition function~\eqref{Z_nm} are chosen among the (A-cyle) contour loops~${\cal A}_a$. This suggests that one can f\/ind~$d_a$ in perturbative power series of $\epsilon$. Regarding $V(z) = O(1)$, we have $W(z) = O(\epsilon)$ and $f(z) =O(\epsilon)$ and the dominant contribution is found in the equation
\begin{gather*}
 4W (z)V'(z) \sim f(z).
\end{gather*}
This is consistent with the expectation that poles of $W (z)$ (zeros of the polynomial~$P(z)$)~\eqref{eq:P-W} are accumulated around the stationary points of the potential $V(z)$. Therefore, we can put $W(z)=\sum\limits_{k \geq 1} \epsilon^k W^{(k)}(z)$ and $f(z)=\sum\limits_{k \geq 1} \epsilon^k f^{(k)}(z)$
and apply the $\epsilon$ expansion to the loop equation~\eqref{loop-classical} directly.

According to the perturbation, the leading order contribution $f^{(1)}(z)$ is related with the f\/illing fraction~$N_k$
\begin{gather}
N_a=\oint_{{\cal A}_a} \frac{f^{(1)}}{2 V'} dz, \label{epsilon^1}
\end{gather}
where ${\cal A}_a$ is the contour encircling around the stationary point $\xi_a$ of the potential. Note that the maximum number of stationary point
is the same as that of~$d_a$. Therefore, $N_a$'s in~\eqref{epsilon^1} f\/ix $d_a$'s to the order of~$\epsilon$. The sub-dominant terms of the f\/illing fraction should vanish, which leads to the null identity. For example, at order of~$\epsilon^2$ one has
\begin{gather*}
0=\oint_{{\cal A}_a} dz \left\{ \frac{f^{(2)}}{2 V'}-\frac{(f^{(1)})'}{4 (V')^2} +\frac{(2 V''-f^{(1)}) f^{(1)}}{8 (V')^3}\right\}.
\end{gather*}

The f\/low equation \eqref{eq:dF} shows that to the lowest order in $\epsilon $
\begin{gather*}
d ( F_N^{(n,m)} ) = 2 \epsilon \sum_{k, a} d c_k U_{ka} \sum_{\ell \ge 1} c_{a+\ell} \left( \sum_{a=1}^{n+m} \frac{ N_a} {\xi_a^\ell}
+ O (\epsilon) \right).
\end{gather*}
Note that $\xi_a(\{c_k\}, \{\hat c_k\}) = 1/ \hat \xi_a(\{\hat c_k\}, \{ c_k\})$. Therefore,
\begin{gather*}
\hat d ( F_N^{(n,m)} ) = 2 \epsilon \sum_{k, a} d \hat c_k \hat U_{ka} \sum_{\ell \ge 1}\hat c_{a+\ell}\left( \sum_{a=1}^{n+m} { N_a} {\xi_a^\ell }
+ O (\epsilon) \right).
\end{gather*}

If one identif\/ies the spectral curve with the Seiberg--Witten curve, then, $x(z)dz $ is the Seiberg--Witten one-form $\lambda$ and the f\/illing fraction of the deformed spectral curve corresponds to the Coulomb branch parameter~$a_k$,
\begin{gather*}
a_a = {\epsilon} N_a+ \epsilon/2.
\end{gather*}
According to the identif\/ication, the partition function has the relation
\begin{gather*}
\frac{ \partial d ( F^{(n,m)} ) }{\partial a_a } = 2 \sum_{k, a} d c_k U_{ka} \sum_{\ell \ge 1} c_{a+\ell} \left( \frac{1} { \xi_a^\ell}
+ O (\epsilon ) \right),
\end{gather*}
where $O(\epsilon)$ corresponds to $O(a_a )$.

\section{Extension of symmetry}\label{sec:4}

\subsection[$W$symmetry]{$\boldsymbol{W}$-symmetry}

$W^{(s+1)}$-symmetry can be easily incorporated if one employs $s$-number of f\/ields. The irregular state is generated by IVO which contains $s$ bosonic f\/ields and its f\/inite number of derivatives. Using the bosonic free f\/ield with holomorphic normalization
\begin{gather*}
\langle \phi^{(a)}(z) \phi^{(b)} (w) \rangle = -\delta^{ab} \log(z-w)
\end{gather*}
one has IVO \cite{PR_IV_2016}
\begin{gather*}
I^{(s|m)}(w)= \exp\left\{\sum_{a=1}^{s}\sum_{k=0}^{m} \frac{c^{(a)}_k}{\hbar k!} \partial_w^k\phi^{(a)}(w) \right\}, \qquad | I^{(s|m)}\rangle = \lim_{w\to0} I^{(s|m)}(w) |0\rangle.
\end{gather*}
Likewise, its adjoint is def\/ined at $1/\zeta$
\begin{gather*}
\hat I^{(s|n)}(\zeta)= R_0^{(s|n)} \exp\left\{\sum_{a=1}^{s}\sum_{\ell=0}^{n }\frac{\hat c^{(a)}_\ell}{\hbar \ell!} \partial_\zeta^\ell~\phi^{(a)}
(1/\zeta)\right\},
\end{gather*}
where $\partial_\zeta$ is the derivative with respect to $\zeta$ and $R^{(s|n)}$ is the scale factor
\begin{gather*}
R_0^{(s|n)} =\exp \left\{ \sum_{a=1}^{s}\sum_{k=0}^{n } \frac{\hat c_0^{(a)}\hat c^{(a)}_\ell}{\ell!}\partial_\zeta^\ell\log(\zeta)\right\}.
\end{gather*}
The adjoint state is def\/ined as
\begin{gather*}
\langle I^{(s|n)}|=\lim_{\zeta \to0}\langle 0 |\hat I^{(s|n)}(\zeta).
\end{gather*}

Their inner product is def\/ined as in Section~\ref{sec:2}
\begin{gather*}
\langle I_0^{(s|n)}| I_0^{(s|m)} \rangle = \lim_{w, \zeta \to 0} \langle 0| \hat I_0^{(s|n)} (\zeta) I_0^{(s|m)}(w) |0 \rangle
=e^{\zeta^{(s)}_{n,m}/\hbar^2 },
\end{gather*}
where $ \zeta^{(s)}_{ n,m}= \sum\limits_{ \ell \ge 1} ^{{\rm min} (m_a,n_a)} \frac{ (\hat {\bf c}_\ell, {\bf c}_\ell)}\ell$ and $\hat {\bf c}_0 + {\bf c}_0 =0$ is assumed. We use bold letters for vectors, ${\bf c}_\ell=\big(c^{(1)}_\ell, \dots, c^{(s)}_\ell \big)$ and $(\hat {\bf c}_\ell, {\bf c}_\ell)$ represents the inner product between two vectors.

Beyond the free theory, we introduce $s$-kind of screening operators $V_{ b {\bf e}_k}$ $(k=1, \dots, s)$
\begin{gather*}
V_{ b {\bf e}_k} = e^{b ( {\bf e}_k , \Phi )} .
\end{gather*}
$\Phi $ is the scalar f\/ields $\Phi=\big(\phi^{(1)}, \dots, \phi^{(s)}\big)$ with $s$ components and $s$-dimensional vector ${\bf e}_k$ are the simple roots of Lie algebra. The bosonic f\/ields with screening operators are represented as Toda f\/ield theory. $({\bf e}_k, \Phi)$ denotes the scalar product. The scalar product $({\bf e}_i, {\bf e}_j)=K_{ij}$ is the Cartan matrix (for $ A_s $ Lie algebra, $K_{ii}=2$, $K_{i i+1}=-1$ and other components are zero). Vertex operator $V_{\alpha} =e^{(\alpha_i, \Phi) }$ has the holomorphic dimension $\Delta_\alpha = (\alpha,(\hat Q- \frac 12 \alpha) )$,
where $\hat Q$ is the background charge vector $\hat Q= Q \rho$ and $\rho$ is the Weyl vector (half of the sum of all positive roots). Therefore, $V_{ b {\bf e}_k} $ has the holomorphic dimension $\Delta_{ b {\bf e}_k}=1$.

Using the screening operators one constructs the inner product of the irregular states in terms of irregular matrix model with $s$-set of variables
\begin{gather*}
\langle I^{(s| n)}| I ^{(s| m)} \rangle_{\bf N } = e^{\zeta_{ n,m}} Z_N^{( {s| n,m})},
\end{gather*}
where $e^{\zeta_{ n,m}}$ is the inner-product due to free f\/ield contribution and $Z_N^{( {s| n,m})}$ is the screening operator contribution
\begin{gather*}
Z_N^{( {s| n,m})} = \left\{ \prod_{a=1}^{s} \prod_{I=1}^{N^{(a)}} \int d\lambda^{(a)}_I \right\} \prod_{a \ge b} \left( \Delta_{ab}\right)^{\beta K_{ab}} e^{ \frac{\sqrt{\beta}}g V^{( {s| n,m})}},
\end{gather*}
where $\Delta_{ab}$ is the Vandermonde determinant
\begin{gather*}
\Delta_{ab}= \delta_{a,b} \prod_{I<J} \big(\lambda_I^{(a)}- \lambda_J^{(a)}\big) +(1-\delta_{a,b}) \prod_{I,J} \big(\lambda_I^{(a)}- \lambda_J^{(b)}\big),
\end{gather*}
and $ V^{(s| n,m)}$ is the potential $V^{( s|n,m)}= \sum_{a} V^{(s| n,m)} (\lambda^{(a)}) $, where
\begin{gather*}
V^{(s| n,m)} \big(\lambda^{(a)}\big) = \sum_{I=1}^{N^{(a)}} \left\{ c^{(a)}_0 \log \big(\lambda^{(a)}_I \big) - \sum_{k=1}^{m^{(a)}} \frac{c_k^{(a )} } {(\lambda^{(a)}_I)^k} -\sum_{\ell=1}^{n^{(a)}} \frac{\hat c_\ell^{(a)} } {\ell} \big(\lambda_I^{(a)}\big)^\ell \right \}.
\end{gather*}
Neutrality condition $\hat {\bf c}_0 + {\bf c}_0 + \epsilon {\bf N } = \mathbf {\epsilon}\rho$ insures that conformal invariance of the system
and symmetry of the inner product. The scaling factor in the adjoint def\/inition is also modif\/ied due to the background charge
\begin{gather*}
R_\epsilon^{(s|n)} =\exp \left\{ \sum_{a=1}^{s} \sum_{k=0}^{n } \frac{(\hat c_0^{(a)} - \epsilon \rho)\hat c^{(a)}_\ell}{\hbar^2 \ell!}
\partial_\zeta^\ell\log \zeta \right\}.
\end{gather*}

In general, one has $(s+1)$-th order spectral curves for the $A_s$ type potential. For example, for $A_2$ case, one has cubic spectral curves
\cite{CR_2015,CRZ_2015, W_2009} if one uses the conformal transformation $\lambda_I^{(1)} \to \lambda_I^{(1)} + \sum_J \frac{\delta}
{(\lambda^{(1)}_I -z)(\lambda_I^{(1)} - \lambda^{(2)}_J)}$ and $\lambda_J^{(2)} \to \lambda_J^{(2)} + \sum_I \frac{\delta} {(\lambda^{(2)}_J -z)(\lambda_I^{(1)} - \lambda^{(2)}_J)}$. Its classical/NS limit is conveniently written in two symmetric forms
\begin{gather}
X_1^3+ \xi_2 X_1+ 3 \epsilon X_1 X_1'+ \epsilon^2 X_1'' = +\frac2{3\sqrt3} \xi_3-\frac\epsilon2 \xi_2' ,\nonumber\\
X_2^3+ \xi_2 X_2+ 3 \epsilon X_2 X_2'+ \epsilon^2 X_2'' =-\frac2{3\sqrt3} \xi_3 -\frac\epsilon2 \xi_2',\label{X2}
\end{gather}
where $X_1$ and $X_2$ are one-point resolvents ($R_1$ and $R_2$) shifted by potential
\begin{align*}
X_1 = 2\left( R_1 + \frac13 (2 V_1'+V_2' )\right),\qquad X_2 =2 \left(R_2 + \frac13(V_1'+2V_2') \right),
\end{align*}
where we use the abbreviation $V_a =V^{( {\bf n,m})}(\lambda^{(a)})$.

There exists also the quadratic form of the spectral curve
\begin{gather}
X_1^2 + X_2^2 - X_1 X_2 + \epsilon (X_1' + X_2') + \xi_2=0, \label{X-quad}
\end{gather}
which presents the Virasoro symmetry. In fact, \eqref{X2} and \eqref{X-quad} are not independent each other. Only two of the three are independent.
$\xi_2(z)$ is the energy momentum tensor (Virasoro current) expectation value
\begin{gather*}
\xi_2(z)=\sum_{k=-2m}^{2n} \frac{A_k}{z^{k+2}} -\sum_{k=-{m}}^{n-1} \frac{d_k}{z^{k+2}} =\frac{ \left\langle{I_m|\hbar^2 T(z)|I_n}\right\rangle}
{ \left\langle{I_m|I_n} \right\rangle} .
\end{gather*}
Here $A_k$ is a constant obtained from the potential with ${\bf c}^{(1)}= {\bf a}$ and ${\bf c}^{(2)}= {\bf b}$
\begin{gather*}
 A_k=2 \epsilon(k+1)( a_k+b_k) -\frac43 \sum_{r+s=k} ( a_r a_s+ b_r b_s + a_r b_s).
\end{gather*}
The mode $d_k$ $( 0\le k \le m-1 )$ is related with the partition function
\begin{gather*}
d_k= v_k \big(F_N^{(m;n)}\big) , \qquad v_k =\sum_{s>0} s \left( a_{s+k} \frac{\partial}{\partial {a_s}} +b_{s+k} \frac{\partial}{\partial {b_s}}
\right)
\end{gather*}
and its dual form $\hat d_k$ $( 0\le k \le n-1 )$ if one replaces ($\bf a, b $) with ($\hat {\bf a}, \hat {\bf b }$).

The expectation of the $\mathcal{W}_3$ current $W(z)$ is given as $\xi_3(z)$
\begin{gather*}
\xi_3(z)=\frac{\left\langle{I^{(n)}|\hbar^2 W(z)|I^{(m)} }\right\rangle} {\langle{I_m|I_n}\rangle} =\sum_{k=-3m }^{3n} \frac{B_k}{ z^{k+3}}
-\sum_{k=-2m}^{2n-1}\frac{e_k}{z^{k+3}},
\end{gather*}
where $B_k$ comes directly from the potential coef\/f\/icients
\begin{gather*}
B_k= \frac4 {3 \sqrt{3}} \sum_{r+s+t=k} \big( 2 (a_r a_s a_t-b_r b_s b_t) + 3 ( a_r a_s b_t-b_r b_s a_t ) \big) \\
\hphantom{B_k=}{} - \frac{\sqrt{3}}2 \epsilon \sum_{r+s=k} \big( 2(k+2 ) (a_r a_s-b_r b_s ) +(r-s) ( a_r b_s- b_r a_s )\big) \\
\hphantom{B_k=}{} +\frac{\sqrt{3}}2 \epsilon^2 (k+1)(k+2) ( a_k-b_k ) .
\end{gather*}
The moment $e_k$ induces the f\/low equation
\begin{gather*}
e_k= \mu_k \big(F_N^{(n,m)} \big),
\end{gather*}
where $\mu_k $ is the $W$-current. To f\/ind the partition function we need the mode with $ n \le k \le 2n-1$ which is
\begin{gather*}
\mu_k = \sum_{\substack{k=r+s-t;\\ t>0}} \sqrt3 {t} \left(-( a_r a_s +2a_r b_s ) \frac{\partial}{\partial a_t} + ( b_r b_s+2a_r b_s ) \frac{\partial}{\partial b_t} \right) .
\end{gather*}
The f\/low equations show the consistent conditions
\begin{gather*}
v_p (d_q) -v_q(d_p) = (p-q) d_{p+q}, \qquad v_p (e_q) -\mu_q(d_p) = (2 p-q) \mu_{p+q}
\end{gather*}
and its duals, replacing ($\bf a, b $) with ($\hat {\bf a}, \hat {\bf b }$), realize the $\mathcal{W}_3$ symmetry.

It is noted in \cite{CRZ_2015} that the spectral curve \eqref{X2} can be put into a third-order dif\/ferential equation of~$\Psi_i(z)$, where $\Psi_i(z)=\exp \big(\frac1\epsilon \int^z X_i(z')dz' \big)$ with $i=1,2$:
\begin{gather*}
\left(\epsilon^3 \frac{\partial^3}{\partial z^3} + \xi_2 \epsilon \frac{\partial}{\partial z} +U_i(z) \right) \Psi_i(z)=0,
\end{gather*}
where $U_1(z)=+\frac2{3\sqrt3} \xi_3-\frac\epsilon2 \xi_2'$ and $U_2(z)=-\frac2{3\sqrt3} \xi_3-\frac\epsilon2 \xi_2'$. Note that $\Psi_i(z)$ corresponds to the normalized expectation value $ \langle I^{(n)}| V_{-{\bf w}_i/b}(z) |I^{(m)} \rangle / \langle I^{(n)}| I^{(m)} \rangle $,
where $ {\bf w}_1 = ( 2 {\bf e}_1 + {\bf e}_2)/3$ and $ {\bf w}_2 = ( 2 {\bf e}_2 + {\bf e}_1)/3$, ($({\bf e}_j, {\bf w}_k) =\delta_{jk}$).
Therefore, the expectation value of $V_{++} (z) =V_{-({\bf w}_1 + {\bf w}_2)/b} (z)$ is given as
\begin{gather*}
\frac{\langle I^{(n)}| V_{++} (z) |I^{(m)} \rangle } {\langle I^{(n)}| I^{(m)} \rangle} = \Psi_1(z) \Psi_2(z).
\end{gather*}

One may also put the spectral curves \eqref{X2} into two coupled third-order dif\/ferential equations of two monic polynomials $P(z)$ and $Q(z)$ of the degree~$N$: $P (z)=\big\langle \prod\limits_{i=1}^{N} (z-x_i )\big\rangle$ and $Q (z)=\big\langle \prod\limits_{j=1}^{M} (z-y_j)\big\rangle $. \eqref{X-quad}~becomes a second-order dif\/ferential equation of two polynomials. Putting $2R_1 (z) = \epsilon P'(z)/P(z)$ and $2R_2 (z) = \epsilon Q'(z)/Q(z)$, one has for~\eqref{X-quad}
\begin{gather*}
\epsilon^2 ( P'' Q - P' Q' + P Q'' ) +2 \epsilon (V_1' P' Q + V_2' P Q')=F P Q
\end{gather*}
and for \eqref{X2}
\begin{gather*} \epsilon^3 P'''+2 \epsilon^2 (2 V_1'+V_2') P'' +\epsilon \big( 4V_1'(V_1'+V_2')+2 \epsilon V_1''-F \big) P'=G_1 P,\\
\epsilon^3 Q'''+2 \epsilon^2 (V_1'+2V_2') Q'' +\epsilon \big( 4V_2'(V_1'+V_2')+2 \epsilon V_2''-F \big) Q'=G_2 Q.
\end{gather*}
$G_1$ and $G_2$ are given in terms of $e_k$ and $d_k$ with irregular poles and zeros
\begin{gather*}
 G_1 = \sum_{k=-2n}^{2m-1} \frac{1}{z^{k+3}} \left\{ -\frac2{3\sqrt3} e_k + \frac23 \sum_{r+s=k} d_r (2 b_s +a_s) \right\}
-\frac{\epsilon}2 \sum_{k=-n}^{m-1} \frac{(k+2) d_k}{z^{k+3}} , \\
 G_2 = \sum_{k=-2n}^{2m-1} \frac{1}{z^{k+3}} \left\{ \frac2{3\sqrt3} e_k + \frac23 \sum_{r+s=k} d_r (2 a_s +b_s) \right\}
-\frac{\epsilon}2 \sum_{k=-n}^{m-1} \frac{(k+2) d_k}{z^{k+3}}.
\end{gather*}

\subsection{Supersymmety}

One may have dif\/ferent type of irregular states if one adopts supersymmetric theory. Considering that $N=1$ super Liouville conformal f\/ield theory is related with the instanton partition function of $N=2$ quiver gauge theories on the ALE space $\mathcal{C}^2/\mathcal{Z}_2$ \cite{BF_2011, BMT_2012}, we expect that the super-symmetric irregular matrix model will provide the useful information on the Argyres--Douglas theory.

The supersymmetric irregular vertex operator is constructed in \cite{PR_SUSY_spectral_2016, PR_SUSY_2016}. We present the operator in the superf\/ield formalism. The super vertex operator $V_\alpha (\omega) = e^{\alpha \Phi} (\omega)$ has holomorphic normalization~\cite{Rash_Stanish_1996}
\begin{gather*}
\left\langle V_{\alpha_1} (\omega_1) V_{\alpha_2} (\omega_2) \right\rangle = (w_1- w_2 -\theta_1 \theta_2)^{-\alpha_1 \alpha_2},
\end{gather*}
where $\Phi$ is the superf\/ield and $\omega=(w, \theta)$ is the holomorphic super-coordinate.

We consider super irregular operator $W^{(q)}(\omega)$ of rank $q$
\begin{gather*}
W^{(q)}(\omega) =\exp \left( \sum_{k=0}^{2q} \frac{\gamma_k}{\hbar } D_\omega^k{ \Phi} (w, \theta )\right),
\end{gather*}
where $D_\omega=\theta\partial_w+\partial_\theta$ is the super-derivative. $\gamma_k $ is a commuting (anti-commuting) number when~$k$ is even (odd). The irregular state lies in NS sector if~$p$ is an integer and in $R$-sector if $p$ is a~half-odd integer. The adjoint operator is def\/ined as
\begin{gather*}
\hat W^{(p)}( \hat \omega) = R^{(p)} \exp \left( {\sum_{k=0}^{2p} \frac{ { \hat \gamma}_k }\hbar D_{ \zeta}^k{ \Phi} (1/\zeta,\hat \theta)}\right),
\end{gather*}
where $\hat \omega$ is another super coordinate $(\zeta, \hat \theta)$ and $R^{(p)}$ is the scale factor
\begin{gather*}
R^{(p)} = \exp \left( \sum_{\ell=0}^{2p} \frac{(\hat \gamma_0-\epsilon) \hat \gamma_\ell}{\hbar^2} D_{\zeta}^\ell \log (\zeta)\right).
\end{gather*}
Using the irregular operator we def\/ine the super irregular state and its adjoint
\begin{gather*}
| W ^{(q)}\rangle =\lim_{w\to0} W^{(q)}(w, \theta) |0\rangle, \qquad \langle W^{(p)}|= \lim_{\zeta \to0}\langle 0 | \hat W^{(p)}(\zeta, \hat \theta).
\end{gather*}

Inner product has the explicit form
\begin{gather*}
\langle W^{(p)}|W^{(q)}\rangle = \lim_{\zeta, w\to 0} \langle 0 | \hat W^{(p)}(\zeta) W^{(q)}(w) |0\rangle
=\exp \left( \frac{ K^{(p;q)}} {\hbar^2 } \right),\\
K^{(p;q)}= \sum_{k \ge 1} \left( \frac{ \hat c_k c_k }k + \hat \xi_k \xi_{k-1} \right),
\end{gather*}
where $\hat \gamma_0 + \gamma_0 =0$ and no background charge ($\epsilon=0$) is assumed. $c_k= (\gamma_{2k} + \gamma_{2k-1}\theta ) k!$ and $\hat c_k= (\hat \gamma_{2k} +\hat \gamma_{2k-1}\hat \theta ) k! $ are commuting. $ \xi_k= ( \gamma_{2k}\theta + \gamma_{2k+1}) k!$ and $ \hat \xi_k= ( \hat \gamma_{2k} \hat \theta +\hat \gamma_{2k+1}) k! $ are anti-commuting. Here we use the super derivative identities $D w^l = l (\theta w^{l-1})$, $D (\theta w^l )= (w^l)$, $D^2 w^l = l (w^{l-1})$, and $D^2( w^l \theta) = l (w^{l-1}\theta )$, $D^3 (\theta w^l )= l (w^{l-1}) $, $D^3 (w^l )= l (l-1)(\theta w^{l-2})$. This identities provide non-vanishing quantities as $w \to 0$: $D^{2n} w^n = n!$, $D^{2n} (\theta w^n) =n! \theta$, $D^{2n-1} (w^n) = n! \theta$ and $D^{2n+1}(\theta w^{n} ) = n!$.

In the presence of background charge we need screening operators $\oint d\zeta ~V_b(\zeta)$ so that neutrality condition $\hat\gamma_0 + \gamma_0 + \hbar Nb =\epsilon$ holds. Then, the inner product has the form
 \begin{gather}
\langle W^{(p)}|W ^{(q)}\rangle_N = \langle W_0^{(p)}|\left( \oint d\zeta V_b(\zeta) \right)^{\!\!N }|W_0^{(q)}\rangle =
e^{ K^{(p;q)}/\hbar^2 } ~Y_{N }^{(p;q)},\nonumber\\
Y_N^{(p;q)} = \int \left [ \prod_{I=1}^N dz_I d\theta_I \right] \prod_{I<J} (z_{IJ} - \theta_I \theta_J )^{\beta}
~e^{\frac{\sqrt{\beta} } {g_s } \sum_{I} V^{(p;q)}(\zeta_I)} .\label{Z_super}
 \end{gather}
Here we use the notation $g_s=i \hbar$. The partition function $Y_N^{(p;q)}$ will be called the irregular super matrix model. The potential $V(\zeta_I)$ of the form $V (\zeta_I) = V_B(z_I) + \theta_IV_F(z_I)$: $ V_B (z)$ and $V_F(z)$ are bosonic and fermionic part of super-potential
\begin{gather*}
V_B(z_I) = c_0 \ln (z_I) - \sum_{k \ge 1} ^q \frac{c_k} {k z_I ^k} - \sum_{k \ge 1}^p \frac{\hat c_k z_I ^k} {k },\qquad
V_F(z_I) = - \sum_{k \ge 0 }^q \frac{\xi_k}{z_I^{k+1}} + \sum_{k \ge 1}^p {\hat \xi_k}{z_I^{k-1}} ,
\end{gather*}
where $ c_0 = \gamma_0$ is used. The potential can be put into more symmetric way if we put $\hat c_k = -c_{-k}$ and $\hat \xi_k = - \xi_{-k}$,
\begin{gather}
V_B(z_I) = c_0 \ln (z_I) - \sum_{k=-p}^ q {}' \frac{c_k} {k z_I ^k},\qquad
V_F(z_I) = - \sum_{k=-p}^ q \frac{\xi_k}{z_I^{k+1}},\label{V_BF}
\end{gather}
where the prime in the summation denotes that no sum on $k= 0$. It is noted that the super irregular matrix model \eqref{Z_super} is also obtained from the colliding limit of the superconformal system with central charge $c= 3/2(1 + 2 Q^2)$.

The loop equation becomes the super-spectral curve \cite{manabe_2016, manabe_2015} for the super-conformal transformation $z_I \to z_I + \frac{ \delta}{z-z_I} $ and $ \theta_I \to \theta_I ( 1 +\frac{ \delta}{2 (z-z_I)^2}) $:
\begin{gather}
x_B (z) x_F (z) + \epsilon x_F '(z) = F_F (z), \nonumber \\
x_B(z)^2 + \epsilon x_B'(z) + x_F(z) V_F' (z) - x_F'(z) V_F(z) = 2F_B (z),\label{spectral_curve-FB}
\end{gather}
where $ x_F (z)$ ($x_B (z)$) is anti-commuting (commuting) one-point resolvent $ \omega_F(z)$ ($ \omega_B(z)$) shifted by potential term,
$x_F(z) = \omega_F(z) - V_F(z) $ ($ x_B(z) = \omega_B(z) + V_B'(z)$)
\begin{gather*}
\omega_B (z) = \epsilon \left\langle \sum_I \frac1{z-z_I} \right\rangle,\qquad \omega_F (z) = \epsilon \left\langle \sum_I \frac {\theta_I}{z-z_I} \right\rangle.
\end{gather*}

 $F_F$ ($F_B$) is also anti-commuting (commuting) holomorphic function
\begin{gather*}
F_F(z) = f_F(z) -V_B'(z) V_F(z) - \epsilon V_F' (z),\qquad
F_B(z) =f_B(z) + \frac12 V_B'^2 + \epsilon V_B'(z)
\end{gather*}
and represent spin 3/2 supercurrent (spin 2 Virasoro) symmetry of the partition function
\begin{gather*}
f_F(z) \equiv \epsilon \left\langle\frac{( V_B'(z) -V_B'(z_I) ) \theta_I - (V_F(z) -V_F(z_I)) } {z-z_I} \right\rangle,\\
f_B(z) = \epsilon \left\langle\sum_I \frac{(V_B'(z) -V_B'(z_I)) + \theta_I (V_F'(z) -V_F'(z_I)) }{z-z_I} +\frac12 \frac{ \theta_I (V_F (z) -V_F (z_I)) } {(z-z_I)^2} \right\rangle.
\end{gather*}
Suppose the potential is given in \eqref{V_BF}. Then, we have $F_F$ ($F_B$) of the form
\begin{gather}
 F_F(z) = \sum_{r=1/2-2p}^{2q-1/2} \frac{ \Omega_{r}} {z^{3/2+r}} + \sum_{r=1/2-p}^{q-1/2} \frac{ \eta_{r}} {z^{3/2+r}} , \qquad
F_B(z) = \sum_{t=-2p}^{2q} \frac{\Lambda_{t}} {z^{2+t}} + \sum_{t=1-p}^{q-1} \frac{d_{t}} {z^{2+t}}, \label{F_FB}
\end{gather}
where $ \Omega_r = \sum\limits_{k+ \ell=r -1/2} c_k \xi_\ell+ \epsilon (r+1/2) \xi_{r-1/2} $ is anti-commuting and $\Omega_{r\ge q+1/2}$ are the eigenvalues of spin 1/2 positive modes. $\Lambda_t = \sum\limits_{k+\ell =t} c_k c_\ell/2 - \epsilon (l+1) c_t /2$ is commuting and $\Lambda_{t \ge q}$ are the eigenvalues of the Virasoro positive modes.

The highest mode eigenvalue $\Lambda_{2q}$ and $ \Omega_{2q-1/2}$ are not vanishing for NS sector. This is because the coef\/f\/icients in the potential can be made of even mode $\gamma_{k}$ ($k$ even) only. The commuting number $c_q= \gamma_{2q}\theta q! $ contains the commuting mode $\gamma_{2q}$. The anti-commuting mode $\xi_{q-1} = \gamma_{2q-2} \theta (q-1)! $ also contains the commuting mode $ \gamma_{2q-2}$. As a result, $\Lambda_{2q}=c_q^2/2 $ and $ \Omega_{2q-1/2}=c_m \xi_{q-1} $ are not vanishing.

However, this is not the case for the Ramond sector where odd modes $\gamma _{k= {\rm odd}}$ are essential ingredients. Especially, $c_m= \gamma_{2q-1}\theta q! $ and $\xi_{q-1} = ( \gamma_{2q-1}) (q-1)! $ contain the odd mode $ \gamma_{2q-1}$ only. As a result the highest positive mode eigenvalues $\Lambda_{2q}=c_q^2/2 $ and $ \Omega_{2q-1/2}=c_q \xi_{q-1}$ should vanish. This is the crucial dif\/ference between the NS sector and R sector.

$\eta_r$ and $d_t$ are expectation values of supercurrent representations $g_r$ and $\ell_t$ (corresponding to right action). $\eta_r$ is anti-commuting
\begin{gather}
\eta_r =g_r \big({-}\hbar^2 \ln Y\big) ,\qquad g_r = \sum_{\ell -k=r} \left( k \xi_{\ell -1/2} \frac{\partial} {\partial c_k}
- c_{\ell } \frac \partial {\partial \xi_{k-1/2}} \right).\label{g_r}
\end{gather}
The relation is obtained if one notices that
\begin{gather*}
\frac{\sqrt \beta}{g} \left \langle \frac{1}{z_I^{k+1}} \right \rangle = k\frac{\partial }{\partial c_{k} } \ln Y ,
\qquad \frac{\sqrt \beta}{g} \left \langle \frac{\theta_I}{z_I^k} \right \rangle = \frac{\partial }{\partial \xi_k } \ln Y.
\end{gather*}
$d_t $ is commuting
\begin{gather}
 d_t= \ell_t \big({-} \hbar^2 \ln Y\big) , \qquad
\ell_t =\sum_{\ell-k=t} \left ( k c_{\ell} \frac{\partial}{\partial c_k}+\left( \frac{ k+\ell }2 \right) \xi_{\ell-1/2 } \frac{\partial}{\partial \xi_{k-1/2}}\right).\label{ell_p}
\end{gather}
The super-f\/low equation \eqref{g_r} and \eqref{ell_p} provides a way to f\/ind the partition function directly from the spectral curve~\eqref{spectral_curve-FB}. The currents~$g_r$ in~\eqref{g_r} and~$l_\ell$ in~\eqref{ell_p} satisfy the commutation relation of right action of the super algebra
\begin{gather*}
[l_\ell, g_r]= \left(r - \frac{\ell}2 \right) g_{r+\ell}, \qquad \{g_r, g_s \} = -2 l_{r+s},\qquad [ l_k , l_\ell ] = -(k-\ell) l_{k+\ell}.
\end{gather*}
This commutation relation provide the consistency condition for the partition function. One may better the derivatives $\partial /\partial c_k$ and
$\partial /\partial \xi_{\ell-1/2}$ in terms of $\ell_k$ and $g_r$. Denoting the derivative elements as $\partial = (\{\partial_m\})=
( \{\frac{\partial}{\partial c_k} \}, \{\frac{\partial}{\partial \xi_{\ell-1/2}} \}) $ and $\zeta = (\{\zeta_m\})= (\{ \ell_k \}, \{ g_r \} ) $
one has $ \partial = {\cal U} \zeta $, where ${\cal U}$ is a $2q \times 2q$ super-matrix. The partition function is given as
\begin{gather*}
d\big({-}\hbar^2 \log Y\big) = d \zeta{\cal U} \tau,
\end{gather*}
where $\tau = (\{\tau_n\} )= (\{ d_k \}, \{ \eta_\ell\})$ in \eqref{F_FB}.

If one considers the degenerate primary superf\/ield $V_{-1/b}$ which has the null state at level~3/2, then the normalized expectation value $\Psi^{(p,q)} (z, \theta)$ of $V_{-1/b}$ is written as following
\begin{gather*}
\Psi^{(p,q)} (z, \theta) =\exp \left(\frac{V^{(p,q)}}{\epsilon } \right)\left\langle \prod (z-\lambda_I - \theta \theta_I) \right\rangle
=\Psi_B^{(p,q)} \Psi_F^{(p,q)},\\
\Psi_B^{(p,q)} (z, \theta) =\exp\left(\frac1\epsilon {\int^z x_B(z') dz' }\right)=P(z) \exp \left(\frac{V^{(p,q)}}{\epsilon } \right),\\
\Psi_F^{(p,q)} (z, \theta) = \exp\left(- \frac{ \theta }\epsilon x_F(z)\right),
\end{gather*}
where $P(z) = \big\langle \prod (z-\lambda_I) \big\rangle$ is the $N$-th order monic polynomial. As in the Virasoro case, the bosonic resolvent is given in terms of the polynomial at the classical/NS limit, $\omega_B(z) = \epsilon P'(z)/P(z)$. The super spectral curve~\eqref{spectral_curve-FB}
reduces to the second-order dif\/ferential equation of the polynomial~$P(z)$ and therefore, the bosonic contribution $\Psi_B^{(p,q)} (z, \theta)$
is given in terms of the polynomial solution. Then, the super-spectral curve~\eqref{spectral_curve-FB} is rewritten as
\begin{gather}
\epsilon \frac{\partial}{\partial z} \big(\Psi_B^{(p,q)} x_F \big) - \Psi_B^{(p,q)} F_F=0,\nonumber\\
\epsilon^2 \frac{\partial^2}{\partial z^2} \Psi_B^{(p,q)} + \left( x_F \frac{\partial V_F }{\partial z} -\frac{\partial x_F}{\partial z} V_F - 2 F_B
\right) \Psi_B^{(p,q)}=0.\label{Psi-diff}
\end{gather}
One thing to note is that the pole structure of bosonic resolvent $\omega_B$ shares with that of the fermionic one~$\omega_F$. This can be seen from \eqref{Psi-diff}. One has the solution of the form $\epsilon x_F(z) = {\tau_F(z)}/{\Psi_B^{(p,q)}(z)}$, where $\tau_F(z)=\int^z \Psi_B^{(p,q)}(z') F_F (z') dz' $. Since $\tau_F(z_\alpha)$ is not zero in general, the obvious conclusion is that the zero $z_\alpha$ of $P(z)$ in $\Psi_B^{(p,q)}(z)$ becomes the pole position of $\omega_F(z)$.

\section{Conclusion}\label{sec:5}

Free f\/ield formalism is a convenient way to construct the irregular states. We provide a few examples; Virasoro states using one boson f\/ield,
W states using multiple boson f\/ields, super-symmetric states with super-f\/ields. Screening operators provide more degrees of freedom and the role of screening operators are given in terms of irregular matrix model. Therefore, the free f\/ield formalism together with the screening operators
demonstrates that the irregular conformal block can be constructed without using the colliding limit.

The partition function is evaluated with the help of the (super) conformal symmetry. The loop equation, or (deformed) spectral curve contains the (super) conformal symmetry. This leads to the f\/low equations of the partition function on the parameter space of the irregular states. In evaluating the partition function, the analytic structure of the spectral curve plays the crucial role.

The spectral curve also provides a way of connecting the irregular matrix model with $N=2$ super gauge theory. The spectral curve is the (deformed) Seiberg--Witten curve of the Hitchin system and the f\/illing fraction (number of screening operator) is identif\/ied with the Coulomb branch parameter of the gauge theory. Therefore, the f\/low equation obtained from the spectral curve ef\/fectively determines the partition function of the Argyres--Douglas gauge theory.

The inner product is the two-point correlation. One may f\/ind multi-point correlations by considering the expectation value of product of regular and/or irregular vertex operators. The expectation value is the product of the free f\/ield contribution and irregular matrix model which is due to the screening operators insertion. The irregular matrix model is simply modif\/ied by adding the potential terms at each position of the inserted regular and/or irregular vertex operators. Therefore, the spectral curve looks the same, only changing the analytic property due to the conformal current for the same type of conformal symmetry: Virasoro symmetry has the quadratic form in $x$, and $W^{(s+1)}$ symmetry has the $s$-th powered form in $x$. The dif\/ference of the irregular singularity from the regular one lies in $\xi_s$'s which represents the conformal symmetry representation. Depending on the type of vertex operators one has the dif\/ferent type of analytic structure in $\xi_s$, regular or irregular poles and zeros. As a result, the f\/low equation changes accordingly and so does the partition function. Depending on the symmetry representation,
one has dif\/ferent analytic structure of the partition function. The understanding of the mathematical structure will be a challenging problem.

\subsection*{Acknowledgements}

The author acknowledges the support of this work by the National Research Foundation of Korea (NRF) grant funded by the Korea government (MSIP) (NRF-2014R1A2A2A01004951). Part of this work presented at ``5th Workshop on Combinatorics of Moduli Spaces, Hurwitz Numbers, and Cohomological Field Theories'', June 6--11, 2016, Moscow, Russia and at the workshop ``Conformal Field Theory, Isomonodromic Tau-Functions and Painlev\'e Equations'',
November 21--25, 2016, Kobe, Japan.

\pdfbookmark[1]{References}{ref}
\LastPageEnding

\end{document}